\documentclass[aps,prl,preprint,12pt]{revtex4}
\usepackage{graphicx}
\usepackage{amssymb}
\usepackage{subfigure}
\usepackage{color}
\usepackage{multirow}
\usepackage{bm}

\begin{document}

\title{Equilibrium and nonequilibrium effects in the collapse of a model polypeptide}
\author{Natalia A. Denesyuk} 
\affiliation{Institute for Physical Science and Technology, University of Maryland, College Park, 
Maryland 20742}
\author{John D. Weeks}
\affiliation{Institute for Physical Science and Technology, University of Maryland, College Park, 
Maryland 20742}
\affiliation{Department of Chemistry and Biochemistry, University of Maryland, College Park, 
Maryland 20742}
\date{\today}

\begin{abstract}
We present results of molecular simulations of a model protein whose hydrophobic collapse proceeds as a 
cascade of downhill transitions between distinct intermediate states. Different intermediates are stabilized by means of 
appropriate harmonic constraints, allowing explicit calculation of the equilibrium free energy landscape. 
Nonequilibrium collapse trajectories are simulated independently and compared to diffusion on the calculated free energy 
surface. We find that collapse generally adheres to this surface, but quantitative agreement is complicated by 
nonequilibrium effects and by dependence of the diffusion coefficient on position on the surface.

\end{abstract}

\maketitle

In Kramers theory a chemical reaction is modeled by Brownian motion of a particle in an effective field of 
force~\cite{Kramers}. The force is related to the change in the free energy of the system as it moves along 
a reaction coordinate that parameterizes the physical or chemical transformation involved. The rate of the reaction is 
then determined by the shape of the (free) energy landscape and by an effective diffusion constant. In recent research the 
energy landscape has been frequently invoked in discussions of complex biological phenomena, such as folding of proteins 
and RNA~\cite{FunnelCombo,DillCombo,RNA}. Furthermore, the energy landscape has been measured experimentally for some 
biophysical systems~\cite{LandscapeCombo}, where it is assumed that a single reaction coordinate captures a process which 
occurs in a high dimensional phase space. In the case of two-state protein folding, theoretical studies have 
indicated that effective diffusion along a reaction coordinate is described by the value of the diffusion coefficient 
which is not constant, as in Kramers theory, but coordinate-dependent~\cite{Chahine,BestHummer}. Such dependence of 
diffusion on the position on the free energy surface will be even more important for downhill protein folding which 
occurs in the absence of a free energy barrier and is, therefore, a diffusion-limited process.

\begin{figure}
\includegraphics[width=8.0cm,clip]{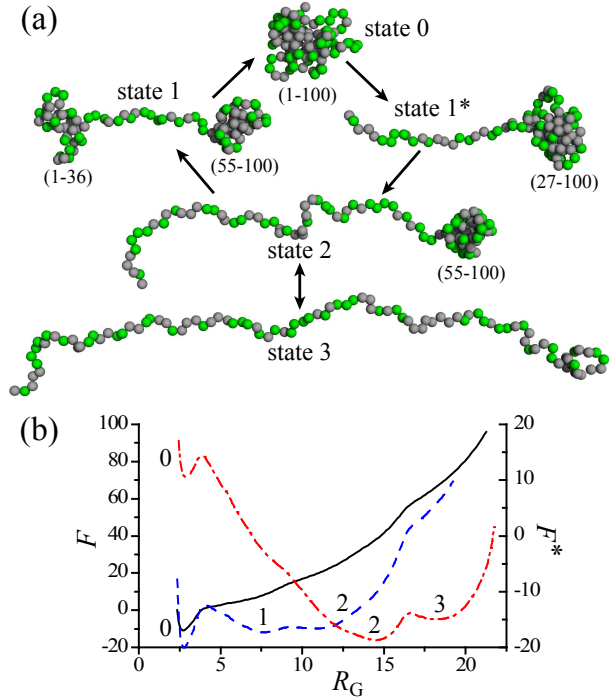}
\caption{\small (a) Polymer conformations corresponding to different states. H-monomers are grey (dark), P-monomers are 
green (light). Numbers in brackets indicate monomers comprising compact globules. (b) Free energies $F$ and $F^*$ for 
the impeding force $f_0=0$ (solid) and $f_0=0.02\varepsilon/\sigma$ (dash), $0.04\varepsilon/\sigma$ (dash-dot). $F$,
$F^*$ are given in the units of $k_{\rm B}T$ and $R_{\rm G}$ in the units of $\sigma$. The dashed-dotted line is shifted 
up by $40k_{\rm B}T$.}
\label{STATES}
\end{figure}

In this letter we study thermodynamics and kinetics of polymer hydrophobic collapse as a basic model for downhill protein 
folding. We find that effective diffusion slows down by an order of magnitude when the polymer enters the free energy 
basin of its collapsed (ground) state, so that subsequent diffusion across this narrow region of the phase space takes an 
appreciable fraction of the overall collapse time. In the absence of a free energy barrier, the application of the 
diffusive model to protein folding requires an accurate knowledge of the entire ground state basin, which cannot be 
deduced from irreversible folding trajectories alone. Below we describe an efficient and accurate method for the 
calculation of multi-dimensional free energy surfaces, including high gradient slopes of the ground state basin. The method 
is based on subjecting a polymer to a set of harmonic constraints and thereby preventing its collapse to the ground state. 
Comparing the results of simulations of an unconstrained polymer and of the polymer harmonically bound to various 
positions on the free energy surface, we conclude that the freely collapsing polymer fails to achieve local equilibrium 
as it enters the ground state basin. Furthermore, for the specific polymer considered here, a local nonequlibrium effect 
is observed at the entrance to the basin, which is caused by a relatively intense collision of different parts of the 
polymer. Such nonequilibrium effects are beyond the scope of the Kramers model and its use may introduce artifacts in 
the numerical analysis of experimentally observed folding times. It remains an open question how 
the Kramers theory can be modified to account for both local and more general nonequilibrium effects. However, in 
computational studies as described here, combined simulations of irreversible folding trajectories and of 
harmonically constrained macromolecules can provide detailed information on their equilibrium free energies as well as 
indicate various nonequilibrium effects.

We consider a simple HP model of a protein, consisting of a polymer chain 
of ``hydrophobic'' (H) and ``hydrophilic'' (P) spherical monomers. HH interactions are modeled by the attractive 
Lennard-Jones potential, $u_{\rm LJ}(r)=4\varepsilon[(\sigma/r)^{12}-(\sigma/r)^{6}]$, where 
$\varepsilon = 5k_{\rm B}T/3$ and $\sigma$ is the monomer diameter. HP and PP interactions are modeled by the repulsive 
part of the same potential. The fraction of P-residues in a sequence is chosen to be equal to the fraction of monomers 
on the surface of a compact spherical globule. This criterion yields 44 P-residues for a chain of length
100~\cite{Endnote1}. The data presented below are for the particular sequence 
H\-P\-H\-H\-H\-P\-P\-P\-P\-H\-P\-H\-P\-H\-H\-H\-P\-P\-H\-H\-P\-P\-H\-H\-P\-P\-H\-H\-P\-H\-H\-H\-P\-H\-H\-H\-P\-P\-P\-H\-P\-H\-P\-H\-P\-P\-P\-P\-H\-P\-H\-P\-P\-P\-H\-H\-H\-P\-H\-P\-P\-H\-H\-H\-H\-H\-H\-H\-H\-P\-H\-P\-P\-H\-P\-P\-H\-P\-H\-H\-P\-H\-H\-P\-P\-H\-H\-P\-H\-H\-H\-H\-H\-P\-H\-H\-P\-P\-H\-H, 
that has been generated randomly to satisfy this condition~\cite{Endnote2}. The polymer bond length $a$ and the angle 
between adjacent bonds $\phi$ are constrained by means of the FENE~\cite{FENE} and harmonic potentials, respectively: 
$u_{\rm F}(a)/k_{\rm B}T=-33.75\ln[1-(2a/3\sigma)^2]$ and $u_{\rm H}(\phi)/k_{\rm B}T=50(\phi-2\pi/3)^2$. The polymer 
dynamics are simulated by solving the Langevin equation for individual monomers $i$, 
$m\ddot{\bm{r}}_i=-\gamma\dot{\bm{r}}_i+\bm{F}_i+\bm{R}_i$, where $m$ is the monomer mass, 
$\gamma$ is the drag coefficient, $\bm{F}_i$ is the conservative force, and $\bm{R}_i$ is the Gaussian random force, 
$\left<\bm{R}_i(t)\bm{R}_j(t^{\prime})\right>=6\gamma k_{\rm B}T\delta(t-t^{\prime})$. The Langevin equation is integrated 
using the leap-frog algorithm with the time step $\Delta t=0.001\tau$ and the drag coefficient $\gamma=m\tau^{-1}$, 
where $\tau=\sigma\sqrt{m/\varepsilon}$ is the unit of time.

The typical collapse pathway of the HP polymer proceeds through states 3, 2, 1, 0, shown in Fig.~\ref{STATES}(a). 
To obtain accurate free energies of these states, we have run a series of simulations with the impeding force 
$\bm{f}_i=f_0(\bm{r}_i-\bm{r}_{\rm{cm}})/R_{\rm G}$ applied to each monomer $i$, where $\bm{r}_i$ is the monomer's 
position and $\bm{r}_{\rm{cm}}$ is the polymer's center of mass. It can be shown that the total work done by this force 
field as the polymer moves from one configuration to another depends only on the difference between the radii of gyration 
$R_{\rm G}$ in the two configurations. This leads to the following relationship between the free energies 
$F(R_{\rm G})$ and $F^*(R_{\rm G},f_0)$ of the free and force-biased polymers,
\begin{eqnarray}
\frac{d F(R_{\rm G})}{d R_{\rm G}}=\frac{\partial F^*(R_{\rm G},f_0)}{\partial R_{\rm G}} + f_0N\nonumber\\ 
= -k_{\rm B}T\frac{\partial \log{M(R_{\rm G},f_0)}}{\partial R_{\rm G}} + f_0N,
\label{free1}
\end{eqnarray}
where $M(R_{\rm G},f_0)$ is the number of times the radius of gyration is found in the range 
$R_{\rm G}\pm\Delta R_{\rm G}/2$ for specific $f_0$, and $N$ is the degree of polymerization. 

In simulations with different $f_0$, successive intermediates in the collapse pathway have been stabilized and extensively 
sampled to yield $F^*(R_{\rm G},f_0)$. For example, at $f_0=0.04\varepsilon/\sigma$, state 2 becomes an equilibrium state 
and 3 is a long-lived metastable state, such that the irreversible transition from 3 to 2 occurs after a wait time of 
$\sim10^6\tau$. At $f_0=0.02\varepsilon/\sigma$, the irreversible transition from state 1 to 0 occurs at a similar time 
scale of $10^6\tau$ and is preceded by frequent, once per $\sim10^4\tau$, reversible transitions between states 2 and 1.

The unbiased free energy $F(R_{\rm G})$ is obtained by averaging the last line in Eq.~(\ref{free1}) over different $f_0$ 
with the weights $w(R_{\rm G},f_0)=M(R_{\rm G},f_0)/\sum_{f_0}M(R_{\rm G},f_0)$ and integrating the result with respect 
to $R_{\rm G}$~\cite{Endnote3}. $F(R_{\rm G})$ is indicated by the solid curve in Fig.~\ref{STATES}(b); the dashed curve
in the same figure is the free energy for $f_0=0.02\varepsilon/\sigma$. Note that at $f_0=0.02\varepsilon/\sigma$ the 
barrier height for the ``unfolding'' transition from state 0 to 1 appears to be $\sim7.7k_{\rm B}T$ which should be 
traversed within time scales of the present simulations. In fact this transition is not observed to occur, which is a clear 
indication that the depth of the ground state basin has been underestimated. 

The spherical globule in Fig.~\ref{STATES}(a) can be forced open by applying a higher value of the pulling force 
$f_0=0.04\varepsilon/\sigma$, for which state 2 is an equilibrium state. During this process, however, a different 
intermediate $1^*$, shown in Fig.~\ref{STATES}(a), is observed in place of 1. When irreversible folding transitions are 
observed (i.e., the reverse transition is observed rarely or not at all in simulation) the apparent free energy drop 
associated with the irreversible transition will depend on the actual time that the molecule spends in its intermediate 
or final state before it proceeds further with its collapse/folding or before the simulation run is terminated. The 
assumption that free energy surfaces can be accurately estimated from irreversible folding trajectories, common in 
simulations of protein folding, leads to systematic underestimation of the depth of the free energy basin of the native 
state. For an accurate assessment of a deep free energy minimum it is necessary to develop a method that uniformly samples 
large gradient slopes leading down to this minimum, which is not achieved by irreversible folding trajectories.  

\begin{figure*}
\includegraphics{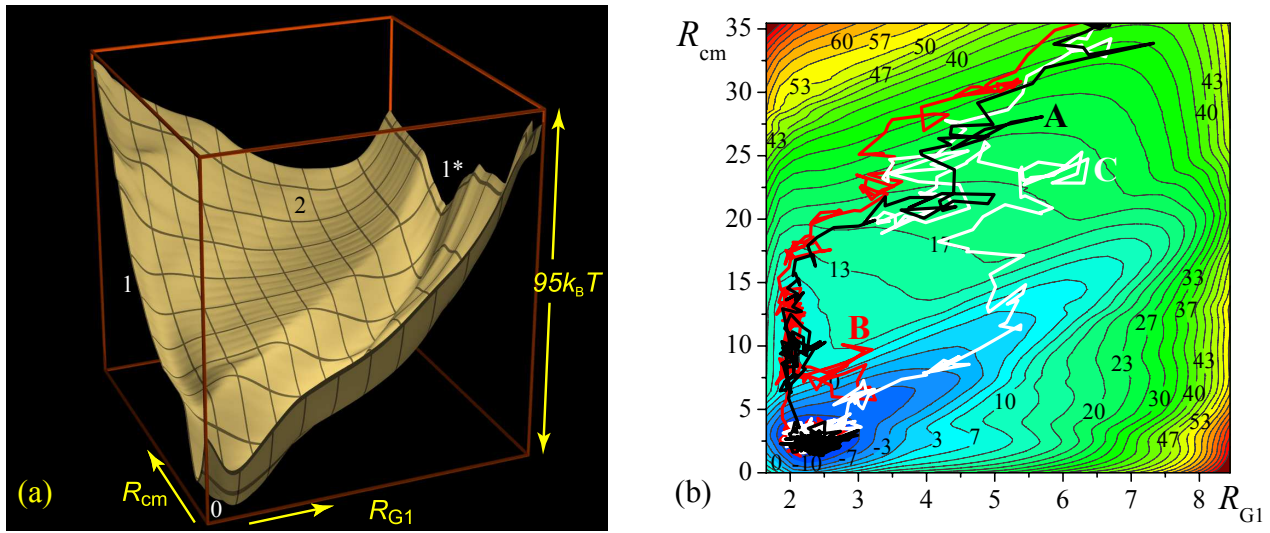}
\caption{\small (a) Free energy surface for polymer collapse from state 2. Two clearly visible trenches belong to 
two distinct pathways connecting states 2 and 0. (b) Contour plot of the same free energy surface together with three 
typical collapse trajectories from state 2. $R_{\rm cm}$ and $R_{\rm G1}$ are introduced in the main text and shown in the 
units of $\sigma$.}
\label{SURFACE}
\end{figure*}

To address this issue, we have completed a new analysis of free energies with two reaction coordinates that distinguish
between states 1 and $1^*$. A choice of reaction coordinates which resolves all on-path and off-path metastable states 
consists of the radius of gyration $R_{\rm G1}$ of the subchain of monomers 1--36 and the distance $R_{\rm cm}$ between the centers 
of mass of subchains 55--100 and 1--12. To systematically investigate the conformational space associated with states 
0, 1, $1^*$ and 2, we have performed numerous simulations in which $R_{\rm cm}$ and $R_{\rm G1}$ are bound to various 
centers $R_{\rm cm}^0$ and $R_{\rm G1}^0$ by harmonic potentials $u_{\rm h1}(R_{\rm cm})=0.5k_1(R_{\rm cm}-R_{\rm cm}^0)^2$
and $u_{\rm h2}(R_{\rm G1})=0.5k_2(R_{\rm G1}-R_{\rm G1}^0)^2$, where $k_1=k_2=0.5\varepsilon/\sigma^2$. In contrast to
the constant force bias employed in Eq.~(\ref{free1}), these harmonic potentials are sufficiently stiff to stabilize 
polymer trajectories on the steep slopes leading down to a deep free energy minimum, which allows an accurate measurement 
of the depth of this minimum and of the transition area. The $R_{\rm cm}$ and $R_{\rm G1}$ derivatives of the unbiased 
free energy $F(R_{\rm cm}, R_{\rm G1})$ in the vicinity of each center $(R_{\rm cm}^0, R_{\rm G1}^0)$ are readily 
calculated from similar derivatives of the biased free energy 
$F^*(R_{\rm cm}, R_{\rm G1}, k_1, R_{\rm cm}^0, k_2, R_{\rm G1}^0)$ by subtracting from the latter 
$k_1(R_{\rm cm}-R_{\rm cm}^0)$ and $k_2(R_{\rm G1}-R_{\rm G1}^0)$, respectively. To establish the partial derivatives over 
the entire phase space we follow the scheme outlined for the 1D case and perform a linear weighted average of all available 
statistics from simulations with different $R_{\rm cm}^0$ and $R_{\rm G1}^0$. A free energy surface is computed 
numerically on a grid by minimizing the sum of squared deviations between its partial derivatives, represented by finite 
differences, and their numerical values obtained from simulations.

The result of this computation for $F(R_{\rm cm}, R_{\rm G1})$ is shown in Fig.~\ref{SURFACE}(a). The narrow trench that 
runs along the $R_{\rm cm}$ axis indicates the collapse transition from state 1 to 0, whereas the wide diagonal trench 
corresponds to the ``unfolding'' transition from state 0 to $1^*$. The accuracy of $F(R_{\rm cm}, R_{\rm G1})$ is 
addressed in auxiliary material~\cite{EPAPS} where we examine the conformational transition between states 1 
and $1^*$. Figure~\ref{SURFACE}(b) shows the contour plot of $F(R_{\rm cm}, R_{\rm G1})$ together with three representative 
trajectories for unconstrained ($k_1=k_2=0$) polymer collapse from state 2. Trajectory A in Fig.~\ref{SURFACE}(b) follows 
the typical collapse pathway $2\to1\to0$. However, 683 out of the total of 10000 simulated collapse trajectories never 
enter the $1\to0$ trench but instead reverse the ``unfolding'' pathway, similarly to trajectory C in Fig.~\ref{SURFACE}(b).

\begin{figure}
\includegraphics[width=8.5cm,clip]{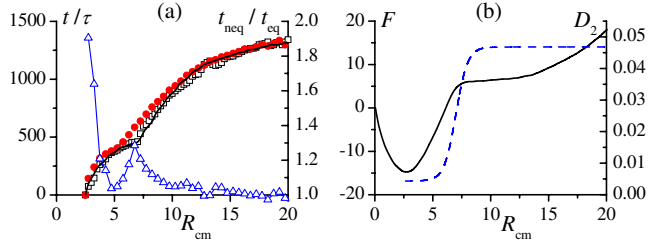}
\caption{\small (a) Collapse times $t_{\rm eq}$ (squares) and $t_{\rm neq}$ (circles) from different initial 
positions in the $1\to0$ trench. The solid curve is the fit of $t_{\rm eq}$ with $D_1=6.4\times10^{-5}$, 
describing lateral vibrations of collapse trajectories in the trench, and $D_2$, shown as dash curve in (b). 
Triangles (right axis) are $t_{\rm neq}/t_{\rm eq}$. In (b), the solid curve is the planar cut $R_{\rm G1} = 2\sigma$ of 
the 2D free energy surface. The units are $k_{\rm B}T$ for $F$, $\sigma$ for $R_{\rm cm}$ and $k_{\rm B}T\gamma^{-1}$ for 
$D_{1}$, $D_{2}$.}
\label{TIMES}
\end{figure}

When collapse proceeds along its typical path, a nonequilibrium effect is occasionally observed at 
the point when blobs 1--36 and 55--100 come into contact. As a result of vigorous mixing caused by the collision of the 
two blobs, monomers 27--36 may break from blob 1--36 and, subsequently, may either return to the same blob, as shown 
by trajectory B in Fig.~\ref{SURFACE}(b), or join with blob 55--100, which appears as a sudden transition from state 1
to $1^*$ (not shown). Aside from this local nonequilibrium effect, we ask if the polymer is generally in quasi-equilibrium 
during its collapse, as assumed in diffusive model. To address this issue, in Fig.~\ref{TIMES}(a) we compare mean 
``equilibrium'' and ``nonequilibrium'' times $t_{\rm eq}$ and $t_{\rm neq}$ of unimpeded polymer collapse from different 
initial positions in the $1\to0$ trench. To compute $t_{\rm eq}$, we have run extensive simulations of the polymer 
bound to various positions in the trench by harmonic constraints $u_{\rm h1}$, $u_{\rm h2}$, that are lifted 
at regular times to initiate the total of 180000 independent collapse trajectories. $t_{\rm neq}$ are derived from other 10000 
trajectories which originate in state 2 and traverse the $1\to0$ trench in passing. The ratio $t_{\rm neq}/t_{\rm eq}$ 
in Fig.~\ref{TIMES}(a) shows a pronounced peak in the transition region between states 1 and 0, indicating delay in 
collapse caused by the local nonequilibrium effect described above. Furthermore, $t_{\rm neq}/t_{\rm eq}$ increases 
monotonically as $R_{\rm cm}\to2.5\sigma$, which suggests that the freely collapsing polymer fails to achieve local 
equilibrium as it traverses steep portions of the reaction pathway.

We also find evidence that different portions of the free energy surface manifest different effective diffusion constants.
Equilibrium times $t_{\rm eq}$ can be fitted to the 2D diffusive model with the diffusion coefficients 
$D_1=6.4\times10^{-5}k_{\rm B}T\gamma^{-1}$ for $R_{\rm G1}$ and the position-dependent $D_2$ for $R_{\rm cm}$, 
shown in  Fig.~\ref{TIMES}(b)~\cite{EPAPS}. Note that $D_2$ decreases by an order of magnitude when the polymer enters 
the free energy basin of its collapsed state, in agreement with the previous work on two-state protein 
folding where a similar slowdown in diffusion is attributed to the increasing polymer compactness~\cite{Chahine}. In 
experiments~\cite{Waldauer} on the folding kinetics of protein L, a ten-fold slowdown in diffusion is observed after the 
initial hydrophobic collapse and attributed to the ruggedness of the energy landscape in the compact unfolded state. In 
the absence of a folding barrier, slow diffusion in the compact unfolded state may contribute significantly to the overall 
folding time, so that details of the energy landscape in this state may become very important. In the present case, a 
planar cut of the free energy surface in Fig.~\ref{TIMES}(b) indicates that the depth of the ground state basin is 
$\sim20k_{\rm B}T$, too large to be computed from irreversible collapse trajectories. Estimating the drag coefficient 
$\gamma$ from Stokes' formula, we find the mean diffusion coefficient $\overline{D_2}$ to be in good agreement with the 
value obtained from experiments on hydrophobic collapse of proteins~\cite{MunozPNAS}. This is not surprising since 
hydrophobic collapse of many heteropolymers involves diffusion and mutual aggregation of locally collapsed 
structures~\cite{LeeThirumalai}, the same elementary processes that take part in the $1\to0$ transition described by $D_2$.

In closing, we can draw some general conclusions from the behavior of this relatively simple HP model. Most applications 
of the diffusive model to protein folding rely on the assumptions that {\it i}) the protein is in local equilibrium 
during its folding, {\it ii}) the energy landscape can be projected onto a single reaction coordinate, 
and {\it iii}) the effective diffusion coefficient is constant, with noted exceptions~\cite{Chahine,BestHummer}. We have 
presented an efficient method for the calculation of multi-dimensional free energy surfaces that has allowed us to 
determine the validity of these assumptions when applied to hydrophobic collapse as a basic model for downhill protein 
folding. We find that polymer collapse generally adheres to the equilibrium energy landscape, although one cannot neglect 
nonequilibrium effects which, in the present case, noticeably delay collapse or cause it to take an altered pathway. 
Nonequilibrium features are not captured in equilibrium free energy calculations and can only be revealed by additional 
analysis of collapse trajectories. At the same time, nonequilibrium effects bias the free energy estimates that are 
obtained from collapse/folding trajectories. Irreversible conformational changes, associated with folding, also interfere 
with accurate determination of free energies from folding trajectories. Furthermore we find, in agreement with previous 
work~\cite{Chahine,Waldauer}, that diffusion slows down by an order of magnitude in the collapsed state, 
resulting from the increasing polymer compactness. In the case of downhill protein folding, this slow diffusion in the 
collapsed state may contribute significantly to the overall folding time. Finally, as illustrated by trajectory A in 
Fig.~\ref{SURFACE}(b), fluctuations in the final state are not necessarily described by the same reaction coordinate as 
collapse to this state, so that projecting trajectories onto a single reaction coordinate may turn out to be inaccurate 
even in the case of a single apparent pathway. In the present study two reaction coordinates have proved sufficient to 
build a thermodynamically and kinetically consistent diffusive model with two distinct, physically meaningful diffusion
coefficients.

This work was supported by NSF through Grant No. CHE-0517818 and through TeraGrid Grant No. TG-CHE070075N.


\begin{thebibliography}{15}
\expandafter\ifx\csname natexlab\endcsname\relax\def\natexlab#1{#1}\fi
\expandafter\ifx\csname bibnamefont\endcsname\relax
  \def\bibnamefont#1{#1}\fi
\expandafter\ifx\csname bibfnamefont\endcsname\relax
  \def\bibfnamefont#1{#1}\fi
\expandafter\ifx\csname citenamefont\endcsname\relax
  \def\citenamefont#1{#1}\fi
\expandafter\ifx\csname url\endcsname\relax
  \def\url#1{\texttt{#1}}\fi
\expandafter\ifx\csname urlprefix\endcsname\relax\def\urlprefix{URL }\fi
\providecommand{\bibinfo}[2]{#2}
\providecommand{\eprint}[2][]{\url{#2}}

\bibitem[{\citenamefont{Kramers}(1940)}]{Kramers}
\bibinfo{author}{\bibfnamefont{H.~A.} \bibnamefont{Kramers}},
  \bibinfo{journal}{Physica} \textbf{\bibinfo{volume}{7}}, \bibinfo{pages}{284}
  (\bibinfo{year}{1940}).

\bibitem[{Fun()}]{FunnelCombo}
\bibinfo{note}{P. G. Wolynes, J. N. Onuchic, and D. Thirumalai, Science {\bf
  267}, 1619 (1995); J. N. Onuchic, Z. Luthey-Schulten, and P. G. Wolynes,
  Annu. Rev. Phys. Chem, {\bf 48}, 545 (1997).}

\bibitem[{Dil()}]{DillCombo}
\bibinfo{note}{H. S. Chan and K. A. Dill, Proteins: Struct. Funct. Genet. {\bf
  30}, 2 (1998); K. A. Dill, Protein Sci. {\bf 8}, 1166 (1999).}

\bibitem[{RNA()}]{RNA}
\bibinfo{note}{D. Thirumalai and S. A. Woodson, Acc. Chem. Res. {\bf 29}, 433
  (1996).}

\bibitem[{Lan()}]{LandscapeCombo}
\bibinfo{note}{M. S. Z. Kellermayer, S. B. Smith, H. L. Granzier, and C.
  Bustamante, Science {\bf 276}, 1112 (1997); J. M. Fernandez and H. Li,
  \textit{ibid.} {\bf 303}, 1674 (2004); M. T. Woodside, P. C. Anthony, W. M.
  Behnke-Parks, K. Larizadeh, D. Herschlag, and S. M. Block, \textit{ibid.}
  {\bf 314}, 1001 (2006)}.

\bibitem[{\citenamefont{Chahine et~al.}(2007)\citenamefont{Chahine, Oliveira,
  Leite, and Wang}}]{Chahine}
\bibinfo{author}{\bibfnamefont{J.}~\bibnamefont{Chahine}},
  \bibinfo{author}{\bibfnamefont{R.~J.} \bibnamefont{Oliveira}},
  \bibinfo{author}{\bibfnamefont{V.~B.~P.} \bibnamefont{Leite}},
  \bibnamefont{and} \bibinfo{author}{\bibfnamefont{J.}~\bibnamefont{Wang}},
  \bibinfo{journal}{Proc. Nat. Acad. Sci. USA} \textbf{\bibinfo{volume}{104}},
  \bibinfo{pages}{14646} (\bibinfo{year}{2007}).

\bibitem[{\citenamefont{Best and Hummer}(2006)}]{BestHummer}
\bibinfo{author}{\bibfnamefont{R.~B.} \bibnamefont{Best}} \bibnamefont{and}
  \bibinfo{author}{\bibfnamefont{G.}~\bibnamefont{Hummer}},
  \bibinfo{journal}{Phys. Rev. Lett.} \textbf{\bibinfo{volume}{96}},
  \bibinfo{pages}{228104} (\bibinfo{year}{2006}).

\bibitem[{End({\natexlab{a}})}]{Endnote1}
\bibinfo{note}{The average content of hydrophilic aminoacids in globular
  proteins is 43.6\%, see T. E. Creighton, \textit{PROTEINS: Structures and
  Molecular Properties} (W. H. Freeman and Company, New York, 1993), 2nd ed.}

\bibitem[{End({\natexlab{b}})}]{Endnote2}
\bibinfo{note}{Similar collapse scenarios were observed for 5 out of 6
  generated sequences, which contained H stretches long enough to act as
  nucleation centers during collapse.}

\bibitem[{\citenamefont{Grest and Kremer}(1986)}]{FENE}
\bibinfo{author}{\bibfnamefont{G.~S.} \bibnamefont{Grest}} \bibnamefont{and}
  \bibinfo{author}{\bibfnamefont{K.}~\bibnamefont{Kremer}},
  \bibinfo{journal}{Phys. Rev. A} \textbf{\bibinfo{volume}{33}},
  \bibinfo{pages}{3628} (\bibinfo{year}{1986}).

\bibitem[{End({\natexlab{c}})}]{Endnote3}
\bibinfo{note}{Differentiating $F^*(R_{\rm G},f_0)$ with respect to $R_{\rm G}$
  eliminates additive constants in these functions that are customarily
  determined by solving a system of nonlinear equations, as in the Weighted
  Histogram Analysis method [A. M. Ferrenberg and R. H. Swendsen, Phys. Rev.
  Lett. {\bf 63}, 1195 (1989)].}

\bibitem[{EPA()}]{EPAPS}
\bibinfo{note}{See EPAPS Document No. [] for additional discussion of the
  $1^*\to1$ transition and for details of the fit of $t_{\rm eq}$ in
  Fig.~\ref{TIMES}(a). For more information on EPAPS, see
  http://www.aip.org/pubservs/epaps.html}.

\bibitem[{\citenamefont{Waldauer et~al.}(2008)}]{Waldauer}
\bibinfo{author}{\bibfnamefont{S.~A.} \bibnamefont{Waldauer}}
  \bibnamefont{et~al.}, \bibinfo{journal}{HFSP J.}
  \textbf{\bibinfo{volume}{2}}, \bibinfo{pages}{388} (\bibinfo{year}{2008}).

\bibitem[{\citenamefont{Sadqi et~al.}(2003)\citenamefont{Sadqi, Lapidus, and
  Mu{\~n}oz}}]{MunozPNAS}
\bibinfo{author}{\bibfnamefont{M.}~\bibnamefont{Sadqi}},
  \bibinfo{author}{\bibfnamefont{L.~J.} \bibnamefont{Lapidus}},
  \bibnamefont{and}
  \bibinfo{author}{\bibfnamefont{V.}~\bibnamefont{Mu{\~n}oz}},
  \bibinfo{journal}{Proc. Nat. Acad. Sci. USA} \textbf{\bibinfo{volume}{100}},
  \bibinfo{pages}{12117} (\bibinfo{year}{2003}).

\bibitem[{\citenamefont{Lee and Thirumalai}(2000)}]{LeeThirumalai}
\bibinfo{author}{\bibfnamefont{N.}~\bibnamefont{Lee}} \bibnamefont{and}
  \bibinfo{author}{\bibfnamefont{D.}~\bibnamefont{Thirumalai}},
  \bibinfo{journal}{J. Chem. Phys.} \textbf{\bibinfo{volume}{113}},
  \bibinfo{pages}{5126} (\bibinfo{year}{2000}).

\end{thebibliography}
\end {document}